\documentclass{article}

\usepackage{subfigure}
\usepackage{mathrsfs}
\usepackage{amssymb,latexsym}
\usepackage{amsmath}
\usepackage{graphicx}
\usepackage{array}
\usepackage{psfrag}
\usepackage{setspace}
\usepackage{verbatim}

\begin{document}
\bibliographystyle{unsrt}

\title{Extending fragment-based free energy calculations with library Monte Carlo simulation: Annealing in interaction space}
\author{Steven Lettieri, Artem B. Mamonov, and Daniel M. Zuckerman \\
\\Department of Computational and Systems Biology\\University of Pittsburgh, Pittsburgh PA,15206}
\maketitle

\begin{abstract}
Pre-calculated libraries of molecular fragment configurations have previously been used as a basis for both equilibrium sampling (via "library-based Monte Carlo") and for obtaining absolute free energies using a polymer-growth formalism.  Here, we combine the two approaches to extend the size of systems for which free energies can be calculated.  We study a series of all-atom poly-alanine systems in a simple dielectric "solvent" and find that precise free energies can be obtained rapidly.  For instance, for 12 residues, less than an hour of single-processor is required.  The combined approach is formally equivalent to the "annealed importance sampling" algorithm; instead of annealing by decreasing temperature, however, interactions among fragments are gradually added as the molecule is "grown."  We discuss implications for future binding affinity calculations in which a ligand is grown into a binding site.
\end{abstract}

\tableofcontents

\newpage
\section{Introduction}
Free energy differences, $\Delta F$, are fundamental to physical chemistry.  In the context of biomacromolecules, $\Delta F$ values can quantify folding stability, relative populations, and binding affinity \cite{Gilson_07_01}.  Although computer simulations have been used to estimate $\Delta F$ values for biomolecules, success has been hampered by well-appreciated sampling problems \cite{Gilson_07_01,Shirts_03_01}.

\noindent A large number of numerical techniques have been used to calculate molecular free energy differences, but a smaller subset is capable of estimating absolute free energies values
\begin{equation}
F = -k_{B}T \;\mathrm{ln Z}
\end{equation}

where $Z$ is the dimensionless configurational partition function.  The most straightforward way to estimate an absolute free energy is using a reference system with an exactly calculable free energy ($F_{\mathrm{ref}}$), so that $F = F_{\mathrm{ref}} + \Delta F$.  $\Delta F$ is then obtained using a standard free energy difference technique, yielding the absolute free energy of the system.  This long-established strategy (e.g., \cite{Frenkel_84_01}) was first suggested for molecular systems by Stoessel and Nowak using a harmonic reference system \cite{Stoessel_90_01}.  Other strategies for calculating absolute free energies are also possible, as demonstrated by the work of Gilson and coworkers \cite{Gilson_07_01, Head_97_01,Chang_04_01,Moghaddam_09_01,Zhou_09_01,Hnizdo_08_01}, as well as by Meirovitch and coworkers \cite{Meirovitch_00_01,White_04_01,Cheluvaraja_04_01,Meirovitch_92_01,Meirovitch_09_01,Meirovitch_10_01,Cheluvaraja_06_01} and Brooks and coworkers \cite{Banba_00_01,Guo_98_01,Kong_96_01}.

\noindent The present study builds on earlier work in our group using reference systems to calculate absolute free energies for molecular fragments which subsequently are combined into a full molecule \cite{Zhang_09_01,Ytreberg_06_01}.  This polymer-growth strategy employs a reference system of non-interacting fragments (e.g. amino acids), which are combined to yield a $\Delta F$ correction accounting for all interactions in the full molecule.  Our earlier studied yielded the absolute free energy for tetra-alanine (Ace-Ala$_{4}$-Nme), but could not easily be applied to significantly larger systems \cite{Zhang_09_01}.  Clark et al.\ developed a closely related fragment-based approach for estimating binding affinities without however, accounting for fragment flexibility \cite{Clark_09_01}

\noindent Here we employ a rigorous ``annealing" strategy \cite{Neal_01_01,Huber_97_01,Lyman_07_02,Lyman_09_01} which integrates the polymer-growth approach with our recently developed library-based Monte Carlo (LMBC) method \cite{Mamonov_09_01,Ding_10_01}.  We anneal not by lowering temperature, but by adding interactions between previously non-interacting fragments.  The addition of interactions is formally equivalent to ``growing" the polymer \cite{Zhang_09_01,Mamonov_09_01}.  Interactions among fragments are added gradually over several stages, permitting the calculation of incremental free energy differences until the full molecule and final free energy is produced.  A weighted configurational ensemble is generated at each stage, which is then ``relaxed" by canonical simulation based on the stage-specific interactions.  This alternation of adding interactions and relaxing is formally equivalent to ``annealed importance sampling" (AIS) \cite{Neal_01_01,Huber_97_01,Lyman_07_02,Lyman_09_01}.  We also employ resampling as a variance-reduction technique, following our previous temperature-based annealing \cite{Lyman_09_01}.

\noindent The annealing strategy is enhanced by our use of LBMC \cite{Mamonov_09_01} for the relaxation phases.  Like other free energy methods, annealed importance sampling employs canonical sampling (here termed ``relaxation") at each stage - which can be performed by any algorithm that correctly samples the stage-specific distributions.  In the present study, LBMC is a natural choice because it is based on fragments and was shown to be highly efficient for sampling flexible peptides in implicit solvent \cite{Ding_10_01}.  The addition of arbitrary interactions for staging is also straightforward in LBMC.  Nevertheless, the sampling method is ``orthogonal" to the free energy calculation, and other canonical sampling algorithms (e.g. \cite{Swendsen,Geyer,Berg,Okamoto,Iftimie,Gelb,Hetenyi,Lyman_06_02,Ytreberg_06_01}) possibly combined with hardware-based improvements \cite{Friedrichs_09_01,Phillips_05_01,Freddolino_08_01,Kale_99_01,Voelz_10_01} could be used instead of LBMC.

\noindent Aside from the issue of calculating absolute free energies, the annealing approach for obtaining $\Delta F$ values can be set in the context of other methods which directly couple equilibrium sampling and free energy estimation \cite{Guo_98_01,Banba_00_01,Neal_01_01,Huber_97_01,Lyman_07_02,Lyman_09_01,Putzer_03_01,Tidor_93_01,Zheng_08_01,Fasnacht_04_01,Li_07_01}.  That is, once the ``staging" of the $\Delta F$ calculation is established --adding interactions in our case, and incrementing a parameter $\lambda$ in many others \cite{Frenkel_84_01} -- equilibrium sampling at each stage can be performed independently or in a coupled way.  Thermodynamic integration \cite{Frenkel_84_01} performs independent simulations for each stage, but in many more recent approaches, ensembles at a given stage are used to aid sampling at other stages \cite{Guo_98_01,Banba_00_01,Neal_01_01,Huber_97_01,Lyman_07_02,Lyman_09_01,Putzer_03_01,Tidor_93_01,Zheng_08_01,Fasnacht_04_01} with ``lambda dynamics" being a good example \cite{Knight_09_01,Kong_96_01,Banba_00_01,Guo_98_01}.  The annealing strategy couples sampling to staging in a uni-directional way, presumably starting from the stage which is easiest to sample.  In our case, it is trivial to generate fully independent configurations for the reference stage of non-interacting fragments; interactions are then ``annealed in."

\noindent The annealing approach combining library-based polymer-growth and LBMC yields good results in this initial study, significantly extending our previous growth work, which was limited to peptides of ~5 residues \cite{Zhang_09_01}.  Using this method, we can compute precise free energies for Ace-Ala$_{12}$-Nme in several hours and for peptides up to 22 residues in about two weeks of computing on a single 3.0 GHz processor core.  Free energies and equilibrium ensembles of smaller peptides can be obtained in seconds or minutes depending on the desired precision.  We validate our results by comparing the equilibrium ensembles produced during annealing to independent Langevin simulations with a collective simulation time of 1 $\mu$s.

\section{Methods}
\noindent We describe how our earlier fragment-based growth method \cite{Zhang_09_01}can be improved using library-based Monte Carlo \cite{Mamonov_09_01,Ding_10_01}.  Formally, however, our procedure is not new, but is a special case of ``annealed importance sampling" (AIS) \cite{Neal_01_01,Lyman_07_02}.  We therefore describe our procedure in terms of AIS, which indeed provides a very natural formal framework.  In our approach however, instead of lowering temperature as in AIS (i.e. annealing in temperature space), we incrementally \emph{add interactions} among molecular fragments. Said another way, interactions between fragments are incrementally "annealed in" - i.e. simply turned on - between successive growth stages.

\subsection{Polymer growth with relaxation: annealed importance sampling}
\noindent The formalism introduced by Neal as AIS \cite{Neal_01_01} will be applied to generalize standard polymer growth algorithms; see also \cite{Huber_97_01}.  It can be described in a straightforward way based on an arbitrary set of un-normalized distributions $\pi_{i}(x)$, with $i=1,2,\ldots, N$ representing the index of the (growth) stage.  In our case, these distributions are standard Boltzmann factors:
\begin{equation}
\pi_{i}(\textbf{x}) \varpropto e^{-U_{i}(\textbf{x})/k_{B}T_{i}}
\end{equation}
\noindent where $U_{i}$ is the potential energy at growth stage $i$ at temperature $T_{i}$.  In our case, annealing is performed only in interaction space so that $T_{i} =298K$ for all $i$.
\noindent Following the previous convention in our growth study \cite{Zhang_09_01}, the initial distribution is $\pi_{1}$, and $\pi_{N}$ is the targeted distribution.  In physical terms, $\pi_{1}$ will represent the distribution of all atoms in non-interacting fragments and $\pi_{N}$ will be the fully interacting molecule.  Full details of the stages are given below in sec. \ref{sec:Staging}.

\noindent Our AIS procedure has only a few simple steps and follows our earlier work \cite{Lyman_09_01}.  The process starts with a well-sampled ensemble of $M$ configurations at stage $i=1$ in the initial distribution $\pi_{1}$.  AIS does not specify a procedure for sampling $\pi_{1}$, but assumes it can be accomplished.  In our case, the non-interacting fragment ensemble can be sampled almost perfectly using internal coordinate Monte Carlo, as described in refs \cite{Zhang_09_01,Mamonov_09_01,Ding_10_01}.

\noindent The ensemble progresses to the next growth stage by ``annealing in" interactions  -- i.e. ``turning on" interactions --  between fragments according to the growth pathway shown in Fig.\ 1 and equations \ref{eq:staging}.  The annealing process shown schematically in Fig.\ 1 corresponds to the case where we are growing a target molecule composed of smaller non-overlapping fragments  $\textbf{x}=(x_{A},x_{B},...,x_{Y})$.

Formally, the $M$ configurations from the current stage $i$ are resampled into the next distribution $\pi_{i+1}$ based on the weights:
\begin{equation}
\label{eq:ratio}
w(\textbf{x})=\pi_{i+1}(\textbf{x})/\pi_{i}(\textbf{x})
\end{equation}

\noindent There are numerous procedures for resampling \cite{Liu_08_01}, but here we use the simplest approach of generating M new configurations for ensemble $i+1$ proportional to the weights from Eq. \ref{eq:ratio}.  This approach leads to some higher-weight configurations being duplicated -- a fact which is exploited in AIS.  After the simple resampling procedure, all weights become equal to one.

\noindent Although the resampled set of $M$ configurations for stage $i+1$ is a statistically valid ensemble for $\pi_{i+1}$, it has suffered some ``attrition" in quality after growth.  Specifically, the uncertainty in calculated observables will be larger than if we had $M$ truly independent configurations -- the non-independence is explicit in the duplicated configurations.  In AIS, one therefore performs some ``relaxation" simulation on each configuration in the ensemble.  This can be done using any canonical sampling algorithm, thus preserving the $\pi_{i+1}$ ensemble but improving the statistical quality.  Our library-based procedure for canonical sampling is described in sec. \ref{sec:LBMC}.  The degree of improvement in ensemble quality depends on the amount of relaxation, a point which we return to later.  Nevertheless, after relaxation, a valid ensemble of M configurations in the $\pi_{i+1}$ ensemble remains.  Reweighting and relaxation are repeated through the growth stages until the targeted distribution, $\pi_{N}$, has been sampled.


\noindent We summarize our AIS procedure as follows:
\begin{enumerate}
\renewcommand{\labelenumi}{(\roman{enumi})}
\item Generate an initial distribution of the $\pi_{1}$ ensemble for stage $i=1$.  This is performed by drawing a random configuration from the pre-calculated library for each fragment: see sec.\ \ref{sec:LBMC}.
\item Resample to the next stage, $i+1$, by ``annealing in" interactions via the weight $w(\textbf{x}) = \pi_{i+1}(\textbf{x})/\pi_{i}(\textbf{x})$. Our stages are specified in sec.\ \ref{sec:Staging}.
\item Relax each configuration via any canonical sampling algorithm, e.g. LBMC, which maintains the $\pi_{i+1}$ distribution.
\item Repeat steps (ii) and (iii) until the target distribution $\pi_{N}(\textbf{x})$ has been reached.
\end{enumerate}

\subsection{Choice of stages: Progressive addition of interactions}
\label{sec:Staging}

\noindent To establish notation, we first divide the full set of coordinates $\textbf{x}$ into N non-overlapping fragments
\begin{equation}
\textbf{x}={x_{A},x_{B},x_{C},...,x_{Y}}
\end{equation}
The total energy, $U(\textbf{x})$, of any fragment-based configuration can be decomposed into two parts.  The first contribution is a sum over the energies internal to each fragment (see $U_{1}$ below); the second is a sum over energies between interacting fragment pairs (see $U_{N}$ below).

\noindent For a target molecule consisting of $N$ fragments, we employ $N$ intermediate models (stages) such that interactions between fragments are gradually turned on along the growth pathway shown in Fig.\ 1.  The first stage, i.e. the reference state $U_{1}$ corresponding to the distribution $\pi_{1}$, is sampled at the library generation stage and only includes interactions internal to each fragment.  Subsequent intermediate stages ``anneal in" the indicated interactions among fragments $A,B,C \ldots Y$.  The energies of the intermediate models can be written recursively as:

\begin{eqnarray}
\label{eq:staging}
U_{1}(\textbf{x}) &=& U_{A}(x_{A})+U_{B}(x_{B})+U_{C}(x_{C})+ \cdots +U_{Y}(x_{Y}) \nonumber \\
U_{2}(\textbf{x}) &=& U_{1}(\textbf{x})+U_{AB}(x_{A},x_{B}) \nonumber \\
U_{3}(\textbf{x}) &=& U_{2}(\textbf{x})+U_{AC}(x_{A},x_{C})+U_{BC}(x_{B},x_{C}) \nonumber \\
\vdots \nonumber \\
U_{N}(\textbf{x}) &=& U_{N-1}(\textbf{x})+ \sum_{\alpha \neq Y} U_{\alpha Y}(x_{\alpha},x_{Y}) \nonumber \\
\end{eqnarray}

\noindent The energy of the last stage, $U_{N}(\textbf{x})$, is the full energy of the desired target molecule.  The sum extends over interactions between the last fragment $Y$, with all previous fragments in the molecule.

\subsection{Free energy calculation in annealed importance sampling}
\noindent The free energy of the fully interacting target ensemble relative to the reference state can be expressed in terms of free energy differences between neighboring levels of the annealing ladder:
\begin{equation}
F_N-F_1=(F_2- F_1)+ (F_3-F_2)+\ldots+(F_N-F_{N-1})
\end{equation}
\noindent or
\begin{equation}
\label{eq:sum}
\Delta F_{1,N}=\Delta F_{1,2}+\Delta F_{2,3}+\ldots+\Delta F_{N-1,N}
\end{equation}

\noindent Possession of the ensembles at each level of the annealing ladder directly permits the calculation of free energy differences between levels.  If the configuration space is progressively narrowed through the stages and the temperature is constant, as in our case\cite{Zhang_09_01}, the required values can be obtained simply using

\begin{equation}
\exp (-\beta \Delta F_{i,i+1}) = \langle \exp (-\beta \Delta U_{i,i+1}) \rangle_{i}
\end{equation}

\noindent where $\Delta U_{i,i+1} = U_{i+1}-U_{i}$ and the ensemble average is over the configurations from stage $i$.

\noindent Although this relation is sufficient for our studies, in more difficult cases, ``two-sided" calculations could be performed - e.g., using the Bennett method \cite{Bennett_67_01}.

\noindent We also point out from Eq. \ref{eq:sum} that if the absolute free energy of stage 1, $F_{1}$, is known then the absolute free energy of stage $N$ can be simply found via
\begin{equation}
F_{N} = F_{1} + \Delta F_{1,N}
\end{equation}

\noindent Absolute free energies of the molecular fragments have already been determined in our previous work \cite{Zhang_09_01}, therefore it is straightforward to convert all free energy differences reported in this paper into their absolute values.

\subsection{Library-based Monte Carlo for relaxation}
\label{sec:LBMC}
AIS requires a canonical sampling procedure for the ``relaxation" process, and we employ library-based Monte Carlo (LBMC)\cite{Mamonov_09_01,Mamonov_10_01}.  LBMC is a natural choice because it can employ the same fragments used in the staging choices, and is also highly efficient for sampling implicitly solvated peptides \cite{Ding_10_01}.  LBMC uses pre-generated libraries of fragment configurations, echoing extensive work with the Rosetta folding program \cite{Rohl_04_01}.  LBMC is a canonical sampling procedure which can be used with an arbitrary forcefield and solvent model.  Full details regarding LBMC have been given in previous work \cite{Mamonov_09_01,Mamonov_10_01,Ding_10_01} but we summarize the essentials here.

In simplest terms, LBMC is an ordinary MC procedure which can employ a special fragment-swap trial move: exchange of the configuration of a fragment with a pre-calculated configuration chosen from a ``library" or ensemble of pre-calculated configurations.  When the library is distributed according to the Boltzmann factor of the target forcefield for all interactions internal to the fragment, the Metropolis criterion\cite{Meteopolis_49_01} is particularly simple:

\begin{equation}
p_{\mathrm{acc}} = \min(1,\exp[-\beta \Delta U^{\mathrm{rest}}])
\end{equation}
\noindent where $\Delta U^{\mathrm{rest}}$ is the change in fragment-fragment interaction energy due to the trial fragment swap.

More precisely, if one is performing a trial swap move by changing a single fragment configuration $x_{J} \rightarrow x_{J'}$, then
\begin{equation}
\Delta U^{\mathrm{rest}} = [ U(x_{A},...,x_{J},...,x_{Y})-U(x_{A},...,x_{J'},...,x_{Y}) ] - [ U_{J'}(x_{J'})-U_{J}(x_{J}) ]
\end{equation}
The first two energy terms are calculated per trial move.  The second two energy terms are simply the energies of the single fragment configurations $J$ and $J'$ -- they are extracted from the pre-calculated libraries.

Many variants of LBMC are possible, but this simple scheme has shown to be successful for flexible all-atom peptides \cite{Ding_10_01}.  In particular, all degrees of freedom are included in the libraries -- so an amino acid fragment consists of all atomic coordinates plus the six connector degrees of freedom which exactly specifiy the position and orientation of the next fragment.  Swap moves are attempted on configurations drawn uniformly from the Boltzmann distributed libraries.

Libraries for Ace, Ala and Nme are generated by internal-coordinate Monte Carlo, as described in our earlier work \cite{Mamonov_09_01}.  Libraries are distributed according to the $T=298\;K$ Boltzmann factor of all OPLSAA energy terms internal to each fragment - both bonded and non-bonded.  The libraries include additional dummy atoms which encode the six degrees of freedom necessary for positioning a fragment with respect to the previous fragment \cite{Mamonov_09_01}.  The fragments Ace, Ala and Nme contain 6, 10, and 6 atoms respectively.  Each fragment library contained $10^{5}$ such distinct configurations and their corresponding energies.  Collectively, these libraries occupy approximately 300MB of computer memory.  Although smaller libraries probably can be effective, we are still investigating optimal sizes.

For the $\pi_{1}$ distribution of non-interacting fragments, LBMC is not necessary.  Rather, the distribution is sampled by drawing a random configuration from the pre-calculated library for each fragment: see $U_{1}$ in Eq. 2.

\subsection{System and simulation details}

We use library-based AIS to calculate free energy changes along the growth pathway in Fig.\ 1 for four polypeptide systems: Ace-Ala$_{4}$-Nme, Ace-Ala$_{12}$-Nme, Ace-Ala$_{16}$-Nme Ace-Ala$_{20}$-Nme.  For all systems under investigation, our libraries implemented the OPLS-AA forcefield \cite{Jorgensen_96_01} with uniform and constant dielectric at constant temperature $298K$.
The uniform dielectric constant was chosen to be $\epsilon = 60$ and no potential cutoffs are used in the calculation of the energy terms.  While other implicit and explicit solvent models are within scope of library-based methods, the goal of this work was to extend library-based free energy calculations\cite{Zhang_09_01} and demonstrate that complete sampling and accurate free energy measurements are easily attainable for larger systems.

During each AIS simulation, the free energy change between growth stages $i$ and $i+1$ is measured according to equation (8) using configurations obtained throughout the relaxation procedure.  For the purpose of examining our data, we also calculate intermediate $\Delta F_{i,i+1}$ values as relaxation proceeds; these values are obtained using Eq. (8) for the set of M partially relaxed configurations at various ``time" points.  The final free energy difference between growth stages $i$ and $i+1$ is then determined by exponentially averaging all intermediate $\Delta F$ values.  The total free energy difference between the target and reference system is calculated by summing the exponentially averaged free energy differences for each growth stage, i.e. via equation (7).

For the systems under investigation, we repeat simulations with various amount of relaxation and ensemble sizes to observe the effects on sampling quality and $\Delta F$ fluctuations.  In principle, these parameters may be adjusted automatically until a desired threshold for free energy accuracy is achieved, however this automation is not implemented in our current work; see sec.\ \ref{sec:improvements}.  The total number of relaxation steps is between $10^{7}-10^{9}$ LBMC trial moves for each simulation and the steps are distributed evenly over each growth stage.  Note that relaxation in the early stages of growth is considerably faster than later stages because there are fewer terms to calculate in Eq.(10).  Specifically, to grow a single polyalanine chain Ace-Ala$_{n}$-Nme containing $n$ alanine residues requires $3.6*10^{n+1}$ energy evaluations.  For each simulation, at least 10 repeats have been performed in order to obtain accurate statistics on variations in sampling quality and free energies $\Delta F_{1,N}$.

\subsection{Validation method}
\label{sec:validationmethod}
To validate proper sampling of our systems, we compare the target ensemble of configurations with those obtained from ten independent Langevin Dynamics trajectories.  Such a comparison is possible by choosing a strongly discriminating representation of phase space based on a Voronoi construction described below.  The ten independent Langevin simulations were performed using TINKER for a collective run time of $1 \mu$s.  All Langevin simulations were run at $\mathrm{T = 298K}$ with a friction constant of 5 $\mathrm{psec^{-1}}$.

We compare the target systems' phase-space distributions with those obtained from well-sampled Langevin Dynamics simulations. Briefly, to obtain a representation of the phase space distribution, we choose five independent and dissimilar reference structures (similarity metric is based on RMSD) from the Langevin Dynamics trajectory of the target system. Configuration space is then partitioned into 5 distinct regions or ``bins" based on a Voronoi procedure so that each bin contains all configurations closest to one reference structure.  This representation of the phase-space distribution provides an extremely sensitive test which is not always ``passed" as seen in the next section.  Full details for this procedure can be found in ref. \cite{Lyman_06_01,Lyman_07_01,Voronoi_07,Zhang_09_01}.

\section{Validation and Results}

\subsection{Sampling Validation}
\label{sec:validation}
First, to check whether sufficient sampling has been performed, Fig. 2 plots the configuration-space distributions mentioned in sec.\ \ref{sec:validationmethod} for the four systems examined.  In each plot, we compare the distributions resulting library-based AIS and Langevin Dynamics simulations.   Error bars have a width of two standard deviations and are based on results obtained from at least 10 independent simulations of each method.  In all cases, there is good agreement between both methods, validating the free energy measurements and sampling capacity of this method.  Note that although the sampling error bars are large for the Ace-Ala$_{20}$-Nme system, the free energy standard deviation is still reasonably small(0.39 kcal/mol) as described below.

For reference, in Fig.\ 3 we plot the configuration-space distributions for the peptide Ace-Ala$_{12}$-Nme as obtained from three methods:  pure LBMC of the full system (no staging), pure growth as in ref. \cite{Zhang_09_01} (no relaxation), and Langevin dynamics simulations.  The data underscores the fact that pure library-based growth simulations are unable to sample these larger systems.  However, by implementing relaxation combined with growth (i.e. AIS), we are able to recover the correct equilibrium distribution -- compare Fig.\ 3 and Fig.\ 2b.

\subsection{Free Energy Measurements and Statistics}
To assess the free energy estimates, we report the mean and standard deviation of the free energy for each polypeptide relative to the non-interacting reference state for varying amounts of relaxation and ensemble sizes as shown in Table 1.  The statistics are based on at least 10 independent simulations for each set of parameters.  The table indicates the computing time required for different levels of precision, although further optimization may be possible (sec.\ \ref{sec:improvements}).

The principal result embodied in Table 1 is that for polyalanine systems up to 16 residues, only a couple of hours is required to reach a level of precision comparable to the accuracy of forcefields ($\sim$0.5 kcal/mol ) \cite{Shirts_03_01}.  It can be seen from Table 1 that free energy variances can be decreased by increasing the overall amount of relaxation.  Importantly, however, increasing the ensemble size implies that each configuration will receive less relaxation if the simulations are to be run in equal amounts of time.  Although the sampling error bars are large for the Ace-Ala$_{20}$-Nme system (e.g Fig.\ 2(d) ), the free energy estimate is still reasonably precise, with a standard deviation of 0.39 kcal/mol.

The series of constant-time simulations for Ace-Ala$_{12}$-Nme in Table 1 indicates that decreasing the ensemble size by a factor is 10 is roughly equivalent to increasing the number of relaxation steps by the same factor insofar as free energy variances are concerned.

A representative plot of intermediate $\Delta F$ measurements during the relaxation of each growth stage for the Ace-Ala$_{12}$-Nme system is shown in Fig.\ 4 along with the exponentially averaged results shown in the inset for each growth stage.  In Fig.\ 4, the sharp decrease in $\Delta F_{i \rightarrow i+1}$ at growth stage $i=6$ is attributed to the fact that, on average, configurations in the ensemble become long enough so that more contacts are formed , e.g. H-bonds and steric clashes.  The LBMC acceptance rate (not shown) also follows a similar trend since there is a larger chance of steric overlap when more atoms are interacting.

We investigated the issue of the choice of ensemble size ($M$) for a given computing investment.  Thus, for Ace-Ala$_{12}$-Nme, four separate runs were performed, each using a total of $10^{8}$ relaxation steps as shown in Table 1.  The standard deviations, $\sigma$,  suggest $M \sim 10^{3}$ is close to optimal for this system.

\section{Discussion}

\subsection{Sampling quality and free energy precision}
What kind of sampling quality is required for reasonable precision (~0.5 kcal/mol) in free energy estimates?  To address this issue, it is interesting to compare results obtained in cases of good and poor sampling.  We have measured the free energy $\Delta F_{1,N}$ using pure growth simulations (no relaxation) for the four polypeptide systems examined, see Table 1.  Pure growth simulations, as mentioned previously, will under-sample configuration space for these larger systems -- see Fig.\ 3 for example.  In the case of the smallest system we examined, Ace-Ala$_{4}$-Nme, relaxation makes very little difference in sampling quality since pure growth simulations are able to accurately predict the correct equilibrium distributions and free energies\cite{Zhang_09_01}.   However, for the larger systems Ace-Ala$_{12}$-Nme and Ace-Ala$_{16}$-Nme, the difference in mean $\Delta F_{1,N}$ values as obtained from poorly-sampled (e.g., Fig.\ 3) and well-sampled simulations(e.g,  Fig.\ 2b) are 1.18 and 1.62 kcal/mol respectively.  This underscores the fact that seemingly small differences in free energies could mask poorly-sampled ensembles.  Such differences would be expected to increase in more complex systems.

The coupling of sampling and free energy calculation suggests one can examine such approaches purely in terms of their ability to provide equilibrium sampling.  To set this in context, note that we recently showed that a fragment-based polymer growth strategy without annealing/relaxation could sample equilibrium ensembles of all-atom peptides but was limited to about five residues \cite{Zhang_09_01}.  The present study, by adding a relaxation phase, greatly extends the size of peptides which can be sampled.

\subsection{Limitations of the relaxed growth method}

The essential limitation of our approach, not surprisingly, is sampling.  It is not a coincidence that our present implementation becomes dramatically more expensive when studying systems beyond the efficient range of current LBMC simulation, roughly 16--20 residues for polyalanine.  Other sampling methods, or improved versions of LBMC, likely could extend the system sizes amenable for free energy estimation.  The basic requirements, as is implicit in Eq. (8) and the subsequent discussion, is to be able to sample the ensembles at every stage to a sufficient degree to permit calculation of the required free energies.  Improved staging, possibly more incremental, could also be useful for larger systems.


\subsection{Possible improvements}
\label{sec:improvements}
Insofar as annealed importance sampling is a formal method based on (i) arbitrary stages and (ii) an arbitrary (correct) sampling method, improvements in either of those components will improve free energy calculations.  Sampling itself could be improved in a number of ways: with LBMC variants better optimized for structured (e.g., folded) systems; with alternative sampling algorithms; or with other hardware optimizations, such as based on graphics processing units (GPUs), for LBMC or for more traditional algorithms \cite{Friedrichs_09_01}.  The relaxation of $M$ configurations at every stage may be amenable to CPU or GPU parallelization.

Staging could also be improved, as is generally the case in free energy calculations \cite{Kofke_97_01,Kofke_98_01}.  Our current procedure adds all interactions for a newly added fragment in a single stage - but this becomes a significant perturbation as system size increases.  The interaction-space staging scheme naturally permits ``sub-staging" of added interactions, and this strategy will be explored in future work.

However, even for a fixed set of stages and a specified sampling algorithm, significant further optimization may be possible.  For instance, as Fig.\ 4 illustrates graphically, some stages are much easier than others - i.e., have greater overlap.  This suggests that the amount of relaxation per stage could be adjusted on the fly based on a convergence criterion, perhaps for a block-averaged variance \cite{Grossfield_09_01} of free energy estimates.  The size of the ensemble,  $M$, could also be optimized in a system and stage-specific way.  These issues will be taken up in future studies.


\subsection{Future applications to protein affinity estimates}
\label{sec:applications}
The resources expended in this study were quite modest, and suggest significantly larger systems could be addressed with more computing.  Yet we do not believe that the present methodology would permit the growth and sampling of all-atom proteins.  What are the prospects for estimating binding affinities?

Importantly, the estimation of binding affinities would not require the growth of a full protein.  Rather, the affinity a ligand for a receptor would be calculated from the difference of free energy values calculated by growing the ligand into solvent and into the receptor binding site.  In other words, only the ligand needs to be grown, and typical ligands are smaller than the peptides studied in this report.  Such a growth procedure adds interactions, and can be seen as the inverse of a decoupling procedure \cite{Hamelberg_04_01,Boresch_03_01}.  At the same time, free energy values require good sampling of the full system, which is never easy for proteins.  In this regard, LBMC was designed to handle hybrid models in a natural way - with an atomistic binding site and a reduced representation elsewhere \cite{Mamonov_10_01}.  Good sampling of such models via LBMC may permit rapid, statistically based affinity estimates within the context of hybrid models.  Such a strategy echoes the work of Roux and coworkers \cite{Beglov_94_01,Wang_06_01,Deng_08_01,Deng_09_01}, but hybrid models would permit potentially important allosteric coupling to regions distant from the binding site.  Another strategy based on fragments has been proposed \cite{Clark_09_01}, but it does not account for flexibility internal to fragments.

\subsection{Improvements in implementation compared to previous work}
By comparison with previous a previous study by our group \cite{Zhang_09_01}, the reported computing times in Table 1 may seem incongruously fast.  Our earlier study examined a series of peptides, with the largest being tetra-alanine, noting that ~50 minutes of computing time were required to obtain a precision of 0.29 kcal/mol for tetra-alanine.  By contrast, in our present work, the same ``pure growth" calculation (no relaxation) required 14 s.  The improvement results primarily from the fact that our previous work was a scripting based post-analysis of data generated by LBMC code, whereas we now calculate free energy values directly within the LBMC code.  Additionally, the growth pathway implemented in this study differs slightly from that in ref. \cite{Zhang_09_01}.  Our current growth pathway (Fig.\ 1) adds new fragments in a more efficient way, increasing overlap between neighboring growth stages.

\section{Summary and Conclusions}
We reported free energy calculations for implicitly solvated polyalanine peptides, ranging in size from four to 20 residues.  The calculations combine two previously developed techniques, a fragment-based polymer growth strategy \cite{Zhang_09_01} and library-based Monte Carlo simulation \cite{Mamonov_09_01,Ding_10_01}.  Our new implementation greatly extends the system sizes amenable to free energy estimation compared to a previous study by our group \cite{Zhang_09_01}.  Because the calculations for the peptides are so inexpensive, we hope the approach can be useful for protein-ligand affinity estimation in the future, as described in sec. \ref{sec:applications}.  The present calculations required seconds to days of single-CPU computing, depending on system size and required precision.

The results here are another application of the memory-intensive strategy of using pre-calculated libraries of molecular-fragment configurations \cite{Rohl_04_01,Mamonov_09_01,Ding_10_01}.  That strategy has been useful for rapid sampling of semi-atomistic protein models \cite{Mamonov_09_01} and of implicitly solvated peptides \cite{Ding_10_01}.  Libraries of configurations have previously been applied extensively in the Rosetta protein folding software \cite{Rohl_04_01} albeit not for canonical sampling.  The potential for ongoing improvements in memory size and access speed, orthogonal to CPU speed, suggests the value of continued pursuit of memory-intensive computations.

From a formal point of view, we have shown that one can perform annealing in interaction space.  That is, our approach is formally equivalent to the annealed importance sampling strategy described by Neal \cite{Neal_01_01}, except that instead of lowering temperature, we add interactions among molecular fragments.


\section{Acknowledgements} We greatly appreciate insightful discussions with Ying Ding, Divesh Bhatt and Xin Zhang, as well as financial support from the NIH (Grants GM076569 and GM070987) and NSF(Grant MCB-0643456).

\newpage
\bibliographystyle{unsrt}
\newpage
\doublespacing
\bibliography{mybib}
\singlespacing

\newpage
\begin{figure}
\begin{center}
  \includegraphics[scale=0.8]{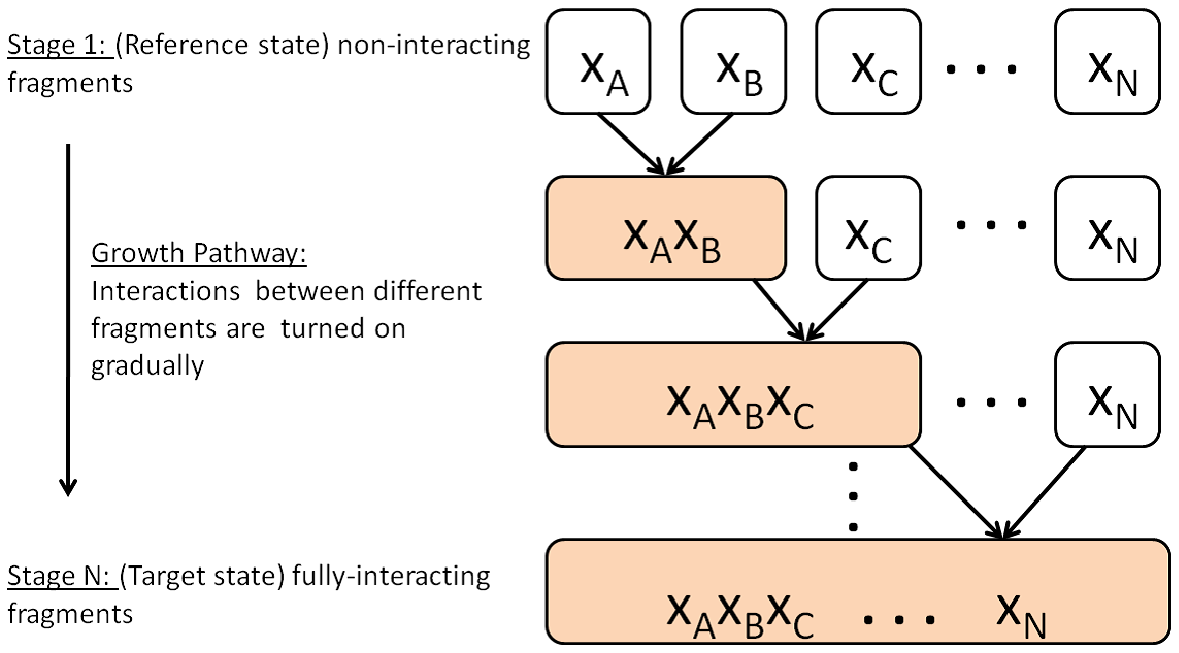}
  \doublespacing
  \caption{A general schematic of the growth and relaxation simulation procedure.   Each box represents an ensemble of fragments.  The process starts with non-interacting fragments (e.g. the amino acids) which are then used to ``grow" the full molecule.  At each intermediate stage, the next fragment in the sequence is added by ``turning on" the additional interactions due to that fragment.  Free energy changes, $\Delta F$,  are computed and accumulated at each growth stage.  This procedure is repeated until the fully interacting molecule is obtained at stage $N$.  The grey colored blocks represent ensembles which have been relaxed according to library-based Monte Carlo.
   }
   \singlespacing
\end{center}
\end{figure}

\newpage
\begin{figure}[htbp]
\doublespacing
  \begin{center}
    \mbox{
      \subfigure[Ace-Ala$_{4}$-Nme]{{\includegraphics[width=60mm,height=40mm]{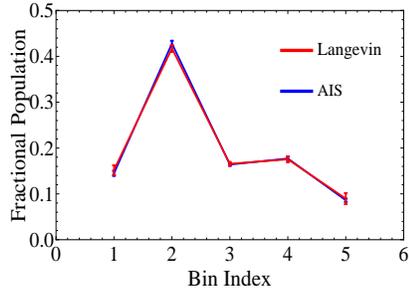}}} \quad
      \subfigure[Ace-Ala$_{12}$-Nme]{{\includegraphics[width=60mm,height=40mm]{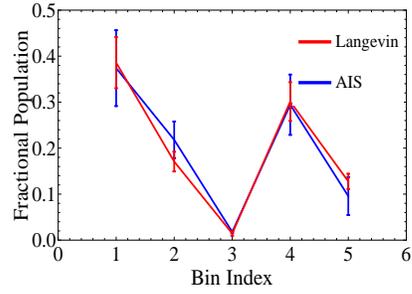}}}
      }
    \mbox{
      \subfigure[Ace-Ala$_{16}$-Nme]{{\includegraphics[width=60mm,height=40mm]{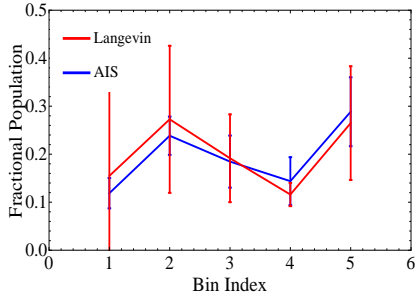}}} \quad
      \subfigure[Ace-Ala$_{20}$-Nme]{{\includegraphics[width=60mm,height=40mm]{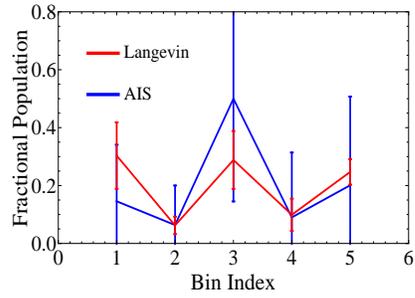}}}
      }
    \caption{Comparison of the equilibrium distributions obtained using both AIS and Langevin Dynamics simulations for the four systems under investigation.  Correct equilibrium sampling has is confirmed for systems a), b) and c) using the new method. System d) has not reached the precision of the 100 nsec Langevin simulations.  Error bars for both of these methods have a width of two standard deviations and are a result of statistics obtained from at least 10 independent simulations.}
    \end{center}
    \singlespacing
\end{figure}

\newpage
\begin{figure}
\doublespacing
\begin{center}
  \includegraphics[scale=0.5]{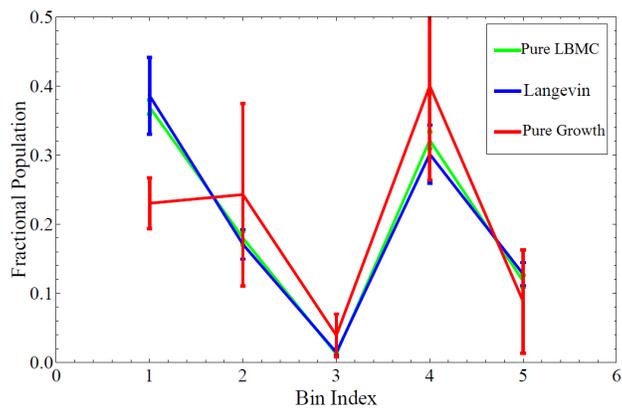}\\
  \caption{Comparison of equilibrium distributions obtained by growth, LBMC and standard Langevin Dynamics simulations.  The plot shows the fractional population in different ``bins" of phase space for the peptide Ace-Ala$_{12}$-Nme.  The error bars have a width of 2 standard deviations and are based on data from 10 independent simulations.  The lines drawn between data points are a guide to the eye.  For a system of this size, a simple growth procedure without relaxation fails to produce the correct distribution.}
\end{center}
\singlespacing
\end{figure}

\newpage
\begin{figure}
\doublespacing
\begin{center}
  \includegraphics[scale=0.40]{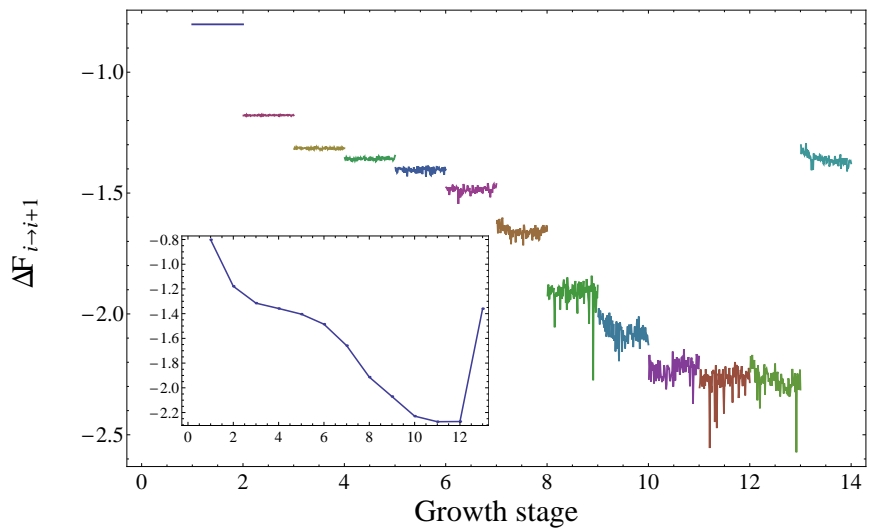}\\
  \caption{Intermediate free energy changes measured during the relaxation process for each growth stage for Ace-Ala$_{12}$-Nme.  Each data point represents Eq. (8) applied to a subset of the data -- different colors are displayed to more clearly distinguish each growth stage.  Shown in the inset is the exponentially averaged final free energy difference that is obtained from averaging all data via. Eq. (8) at each growth stage.  Lines between data points are drawn to guide the eye.  These are representative results from a single simulation.}
\end{center}
\singlespacing
\end{figure}

\newpage
\begin{table}[]
\doublespacing
\caption{Free energy statistics.  The mean values ($\mu$) and standard deviations ($\sigma$) in kcal/mol for the four peptide systems are tabulated as a function of ensemble size and amount of relaxation.  Statistics are based on at least ten independent simulations.}
\singlespacing
\begin{center}
\scalebox{0.8}{
\begin{tabular}{|c|c|c|c|c|c|c|}
\hline
\centering
System & Ensemble Size ($M$) & Tot no. Steps &  $\mu(\Delta F_{1,N})$ & $\sigma(\Delta F_{1,N})$ & CPU time \\ [0.5ex]
\hline\hline
Ace-Ala$_{4}$-Nme & $10^{3}$ & no relaxation/pure growth & $-5.581$  & 0.043 & 14 sec.\\
                  & $10^{3}$ & $10^{7}$ & $-5.562$  & 0.001 & 7.8 min.\\
\hline
Ace-Ala$_{12}$-Nme &$10^{4}$ & no relaxation/pure growth & $-20.19$  & 0.985 & 20 sec.\\
                    & $10^{1}$ & $10^{7}$ & $-21.22$  & 0.299 & 1 hr.\\
                     & $10^{1}$ & $10^{8}$ & $-21.21$  & 0.235 & 8.6 hrs.\\
                  & $10^{2}$ & $10^{8}$ & $-21.38$  & 0.223 & 8.6 hrs.\\
                  & $10^{3}$ & $10^{8}$ & $-21.24$  & 0.169 & 8.6 hrs.\\
                  & $10^{4}$ & $10^{8}$ & $-21.28$  & 0.191 & 8.6 hrs.\\
                   & $10^{4}$ & $10^{9}$ & $-21.37$  & 0.057 & 3.5 days\\
\hline
Ace-Ala$_{16}$-Nme & $10^{4}$ & no relaxation/pure growth  & $-29.78$  & 1.191 & 35 sec. \\
                    & $10^{1}$ & $10^{7}$ & $-30.78$  & 0.459 & 1.9 hrs. \\
                   & $10^{2}$ & $10^{8}$ & $-31.02$  & 0.396 & 16.6 hrs. \\

                   & $10^{4}$ & $10^{9}$ & $-31.40$  & 0.179 & 7 days \\
\hline
Ace-Ala$_{20}$-Nme & $10^{3}$ & no relaxation/pure growth &$-38.99$  & 2.101 & 13 sec. \\
                    & $10^{1}$ & $10^{8}$ &$-41.66$  & $0.870$ & $1.3$ days \\
                    &$10^{3}$ & $10^{9}$ &$-42.45$  & $0.394$ & $14$ days \\ [1ex]
\hline
\end{tabular}}
\end{center}
\end{table}

\end{document}